\documentclass[12pt,a4paper,english]{article}
\usepackage[T1]{fontenc}
\usepackage[utf8]{inputenc}
\usepackage{float}
\usepackage{graphicx}
\usepackage{xcolor}
\usepackage[normalem]{ulem}

\makeatletter

\providecommand{\tabularnewline}{\\}

\newcommand{\lyxaddress}[1]{
	\par {\raggedright #1
	\vspace{1.4em}
	\noindent\par}
}

\newcommand{\diff}{\mathrm{d}}
\makeatother

\usepackage{babel}

\begin{document}
\title{Observational signatures of misaligned double-ring and double-torus configurations around a Schwarzschild black hole}
\author{Dmitriy Ovchinnikov, Jan Schee, and Zdeněk Stuchlík}
\maketitle

\lyxaddress{Research Centre for Theoretical Physics and Astrophysics, Institute of Physics, Silesian University in Opava, Bezru\v{c}ovo n\'{a}m. 13, Opava, Czechia}
\begin{abstract}
{We investigate the observational signatures of an idealized
double-ring and double-torus system orbiting a Schwarzschild black hole,
allowing the two emitting components to have mutually inclined symmetry
axes.} {Using general-relativistic ray tracing, we construct
frequency-shift maps, bolometric flux maps on the observer's screen, and
the corresponding spectral line profiles of the emitted radiation.} 
{The single equatorial torus is used as a reference configuration
in order to isolate the effect of the second emitting component and of the
mutual misalignment of the two structures.} 
{We show that the presence of two non-coplanar emitting
structures produces characteristic multi-peak spectral profiles and
asymmetric bolometric-flux distributions. These signatures are imprinted
both in the line-profile morphology and in the $\alpha$-profiles of the
bolometric flux, providing simple diagnostic features of non-coplanar
multi-component accretion structures.}

\end{abstract}

\section*{Introduction}

{
Recent progress in very-long-baseline interferometry has opened a direct
observational window onto the horizon-scale environment of supermassive
black holes. The Event Horizon Telescope (EHT) observations of M87* and
Sgr A* have revealed compact, asymmetric emission regions shaped by strong
gravitational lensing, relativistic plasma motion, and radiative processes
in the immediate vicinity of the central object
\cite{EHT:2019:M87I,EHT:2022:SgrAI}. These observations are commonly
interpreted with the aid of general-relativistic magnetohydrodynamical
(GRMHD) simulations combined with general-relativistic radiative transfer
and ray-tracing techniques
\cite{EHT:2019:M87V,EHT:2022:SgrAV,MoscibrodzkaGammie:2018:MNRAS}.
Such simulations provide the most realistic description of black-hole
accretion flows currently available. At the same time, simplified
semi-analytical models remain useful, because they isolate individual
geometrical and relativistic effects and allow one to identify which
features of images and spectra are caused by a small number of physical
parameters.
}

{
Accretion onto compact objects is one of the main mechanisms responsible
for strong X-ray emission from black-hole systems
\cite{ShakuraSunyaev:1973:AA,NovikovThorne:1973,PringleRees:1972:AA}.
The observed radiation carries information about the geometry, motion, and
optical properties of the emitting plasma, since relativistic Doppler
shifts, gravitational redshift, light bending, and focusing strongly modify
the observed spectra and line profiles
\cite{Cunningham:1975:ApJ,Fabianetal:1989:MNRAS,Laor:1991:ApJ}.
In many theoretical models the emitting matter is represented either by
geometrically thin disks
\cite{ShakuraSunyaev:1973:AA,NovikovThorne:1973},
by narrow radiating rings
\cite{KarasSochora:2010:ApJ,Sochoraetal:2011:MNRAS},
or by geometrically thick, pressure-supported tori
\cite{FishboneMoncrief:1976:ApJ,Abramowiczetal:1978:AA,Kozlowskietal:1978:AA}.
In particular, thick tori provide a useful idealization of non-spherical
accretion structures and have been widely used to study relativistic
images, radiation fluxes, and spectral line profiles in the strong-field
region of black holes
\cite{Abramowiczetal:1978:AA,WuWang:2007:MNRAS}.
}

{
A further motivation for considering multi-component emitting structures
comes from the theory of ringed accretion disks. In this framework, the
accretion environment may consist of several pressure-supported tori or
ring-like structures orbiting the same central black hole
\cite{PugStu:2017:ApJS, PugStu:2015:ApJSS}. Such configurations
can be interpreted as idealized stages of a more complicated accretion
history, for example when matter with different angular momentum is
supplied to the central region. The dynamical evolution of interacting
double-torus systems has also been studied numerically, showing that such
systems are generally not expected to remain permanently isolated and may
undergo interaction, accretion, or merging phases \cite{Baretal:2022:ApJ}.
}

{
The possibility of non-coplanar accretion structures is motivated by a
broader class of tilted, warped, or broken accretion flows. Misaligned
accretion disks can arise when the angular-momentum axis of the inflowing
matter does not coincide with a preferred symmetry axis of the system.
Numerical simulations of tilted accretion flows have shown that their
images and spectral line profiles can differ substantially from those of
aligned disks and may depend strongly on the observer's azimuthal position
\cite{DexterFragile:2011:ApJ}. In addition, disk tearing mechanisms can
produce several distinct ring-like components with different orbital planes
\cite{Nixonetal:2012:ApJL,Nealonetal:2015:MNRAS}. Although such effects are
often discussed in connection with rotating black holes, the Schwarzschild
spacetime provides a clean reference case in which the influence of the
source geometry can be studied without frame dragging.
}

{
Misaligned aggregates of pressure-supported tori around static
Schwarzschild black holes have recently been investigated from the point of
view of hydrodynamical constraints, possible collisions, limiting
configurations, energetics, and stability properties
\cite{PugStu:2020:MNRAS,PugStu:2020:CQG}. These studies motivate the
question addressed here from a complementary point of view: how would a
simplified non-coplanar double structure appear to a distant observer?
}

{
In the present work we study the ray-traced observational signatures of an
idealized double-ring and double-torus system in Schwarzschild spacetime.
The two emitting components are allowed to have mutually inclined symmetry
axes. We do not assume that such a configuration represents a permanently
stable global equilibrium. Rather, it is treated as a stationary radiative
snapshot of a possible transient or quasi-stationary stage of a
multi-component accretion flow. For this system we compute frequency-shift
maps, bolometric flux maps on the observer's screen, $\alpha$-profiles of
the bolometric flux, and spectral line profiles. By comparing the
non-coplanar double configurations with the corresponding single-torus and
coplanar cases, we identify characteristic signatures produced by the
presence of a second inclined emitting component.
}

\section{Model}

{
We consider two idealized models of radiating structures orbiting a
Schwarzschild black hole. The first model consists of two narrow
Keplerian rings and is used to study the formation of spectral line
profiles in a simple multi-component source. The second model replaces
the rings by finite-thickness pressure-supported tori and is used to
construct bolometric flux maps and the corresponding one-dimensional
flux profiles on the observer's screen. Throughout the paper we use
geometrized units \(G=c=M=1\). The details of the null geodesic equations,
the observer's impact parameters, the rotated coordinate frame, and the
redshift calculation are summarized in Appendix~\ref{app:raytracing}.
}

\subsection*{Rings}

{
A radiating ring is modeled as a narrow circular distribution of test
particles moving along Keplerian circular geodesics. The ring radius is
restricted to the region outside the innermost stable circular orbit,
\(r_{\rm ISCO}=6M\). The corresponding Keplerian angular velocity is
}

\begin{equation}
    \Omega_K=\frac{1}{r^{3/2}}.
\end{equation}

{
The local emitted specific intensity is modeled by a Gaussian line profile,
}

\begin{equation}
    I_e(\nu_e)=I_0(r)\exp\left[-\frac{(\nu_e-\nu_0)^2}{\sigma^2}\right]
\end{equation}

{
where \(\nu_e\) is the emitted frequency, \(\nu_0\) is the rest-frame line
frequency, and \(\sigma\) determines the intrinsic width of the emitted
line. The radial dependence of the emissivity normalization \(I_0(r)\) is
specified for the radii used in the simulations in Table~\ref{tab:I0}.
}

\begin{table}[H]
    \centering
\caption{{Values of the emissivity normalization \(I_0(r)\) used
    for the ring simulations.}}
\label{tab:I0}
    \begin{tabular}{c|c}
         $r$ & $I_0$\\
         \hline
         $6$ & $1.00$\\
         $8$ & $0.02$\\
         $10$ & $0.01$\end{tabular}    
\end{table}

\begin{figure}
    \centering
    \includegraphics{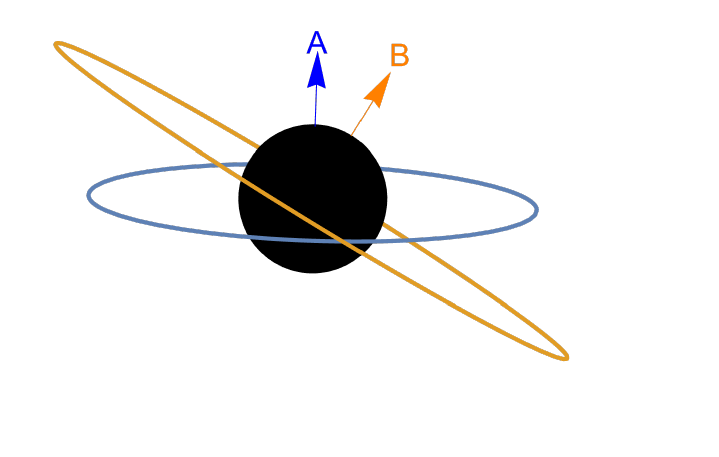}
    \caption{{Schematic representation of the double-ring radiating
    structure. The rings \(A\) and \(B\) may have mutually inclined orbital
    planes.}}
    \label{fig:rings}
\end{figure}

{
The double-ring system consists of an inner ring \(A\) and an outer ring
\(B\), as shown schematically in Fig.~\ref{fig:rings}. Because the
Schwarzschild spacetime is spherically symmetric, each ring can be assigned
its own orbital plane. The orientation of each ring is specified by the
normal vector to its orbital plane. In general, the normal vectors
\(\vec n_A\) and \(\vec n_B\) are not parallel, and the angle between them
defines the mutual misalignment of the two rings. A schematic representation of this
configuration is shown in Fig.~\ref{fig:rings}.
}

{
The null geodesics connecting the emitting ring to the observer are
obtained from the separated geodesic equations in Schwarzschild spacetime.
Each geodesic is uniquely characterized by the impact parameters \(l\) and
\(q\), which are related to the observer's screen coordinates
\((\alpha,\beta)\). In the ring calculations we include the primary-image
contributions specified in Appendix~\ref{app:null-geodesics}. The observed
specific flux profile of a narrow ring is then written as
}
\begin{equation}
    F(\nu_o;r_e)
    =
    \Delta r_e
    \int\limits_{g_{\min}}^{g_{\max}}
    g^3
    I_e(\nu_o/g;r_e)
    \left|
    \frac{\partial(\alpha,\beta)}
    {\partial(r_e,g)}
    \right|
    \diff g ,
    \label{eq:ring-observed-flux}
\end{equation}

{
where \(g=\nu_o/\nu_e\) is the frequency-shift factor. The factor \(g^3\)
follows from the invariance of \(I_\nu/\nu^3\), while the Jacobian of the
mapping \((\alpha,\beta)\rightarrow(r_e,g)\) is evaluated numerically, as
described in Appendix~\ref{app:ring-line}.
}

\subsection*{Tori}

{
We next consider finite-thickness radiating structures. Each torus is
modeled as a stationary, axisymmetric, pressure-supported perfect-fluid
configuration in Schwarzschild spacetime. The model corresponds to the
standard Polish-doughnut construction with constant specific angular
momentum \(l_{\rm disk}\). For such a configuration the equipressure and
equidensity surfaces coincide with the equipotential surfaces
\cite{Abramowiczetal:1978:AA}
}

\begin{equation}
W_{0}
=
W(r,\theta')
=
\frac{1}{2}
\ln
\frac{
(r-2M)r^{2}\sin^{2}\theta'
}{
r^{3}\sin^{2}\theta'
-
(r^{2}-2M)l_{\rm disk}^{2}
}.
\label{eq:torus-potential}
\end{equation}

\begin{figure}[H]
\begin{centering}
\begin{tabular}{ccc}
\includegraphics[scale=0.3]{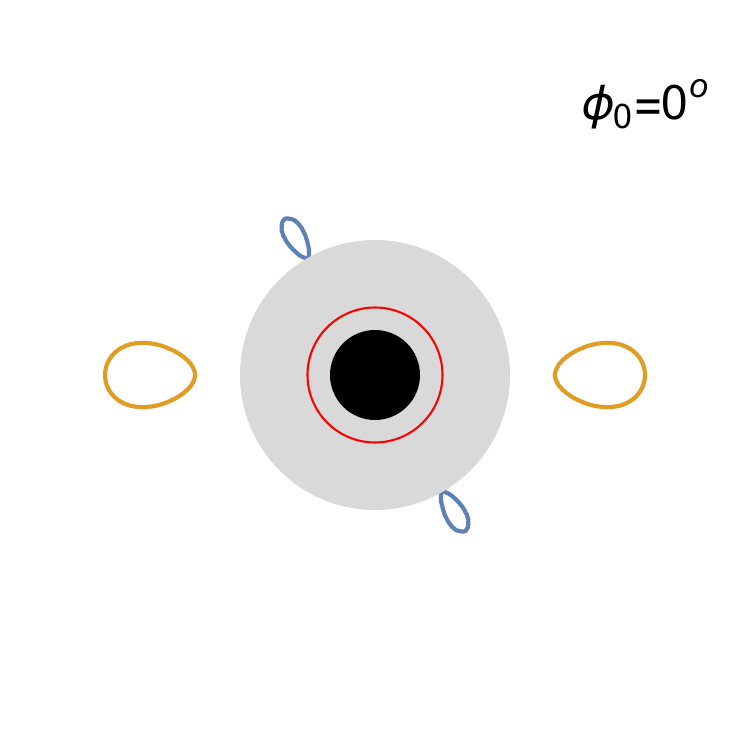} & \includegraphics[scale=0.3]{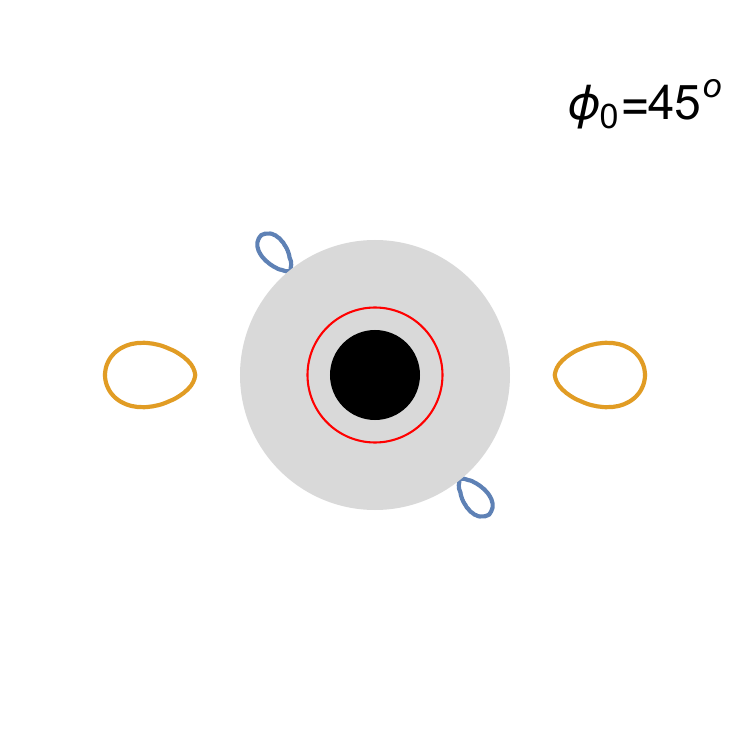} & \includegraphics[scale=0.3]{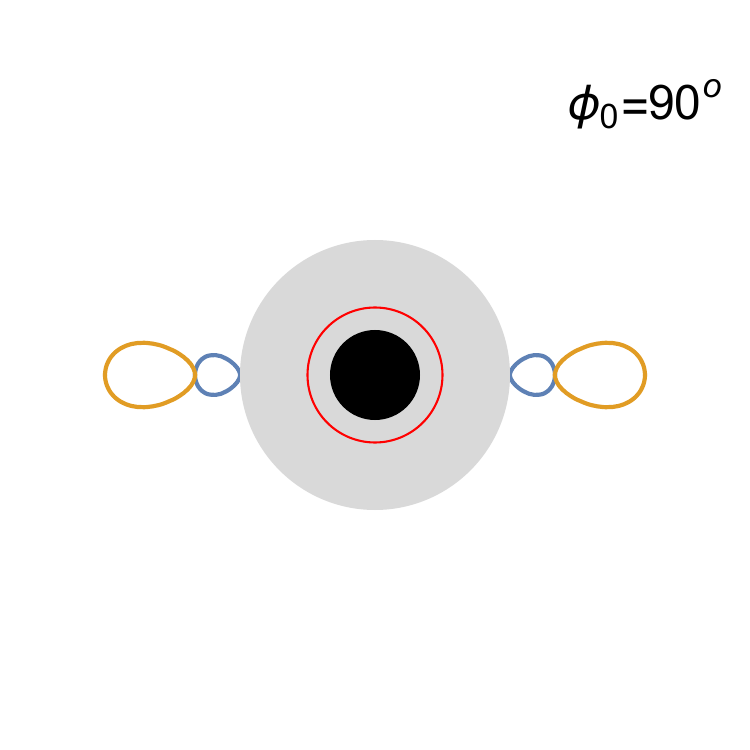}\tabularnewline
\end{tabular}
\par\end{centering}
\caption{{Meridional cross-sections of a double-torus structure
viewed at three representative azimuthal angles
\(\phi_o=0^\circ,45^\circ,90^\circ\). The yellow and blue contours denote
the outer edges of the two tori. The shaded region corresponds to the
sphere \(r_{\rm ISCO}=6M\), the red circle marks the photon sphere
\(r_{\rm ph}=3M\), and the black disk represents the black-hole horizon
at \(r_h=2M\).}
\label{fig:The-meridional-crosssection}}
\end{figure}

{
Here \(\theta'\) is the polar angle measured in the coordinate frame
adapted to the symmetry axis of the corresponding torus. For an equatorial
torus, \(\theta'=\theta\). For an inclined torus, the relation between
\((r,\theta,\phi)\) and \((r',\theta',\phi')\) is obtained by a rotation of
the source-adapted frame about the original \(y\)-axis; the explicit
transformation is given in Appendix~\ref{app:rotated-frame}.
The double-torus configuration is obtained by combining two such
equipotential surfaces with different radial extensions and, in general,
different symmetry axes. Representative meridional cross-sections of this
geometry, viewed at different azimuthal angles, are shown in
Fig.~\ref{fig:The-meridional-crosssection}.
}

\begin{figure}[H]
\begin{centering}
\begin{tabular}{cc}
\includegraphics[scale=0.3]{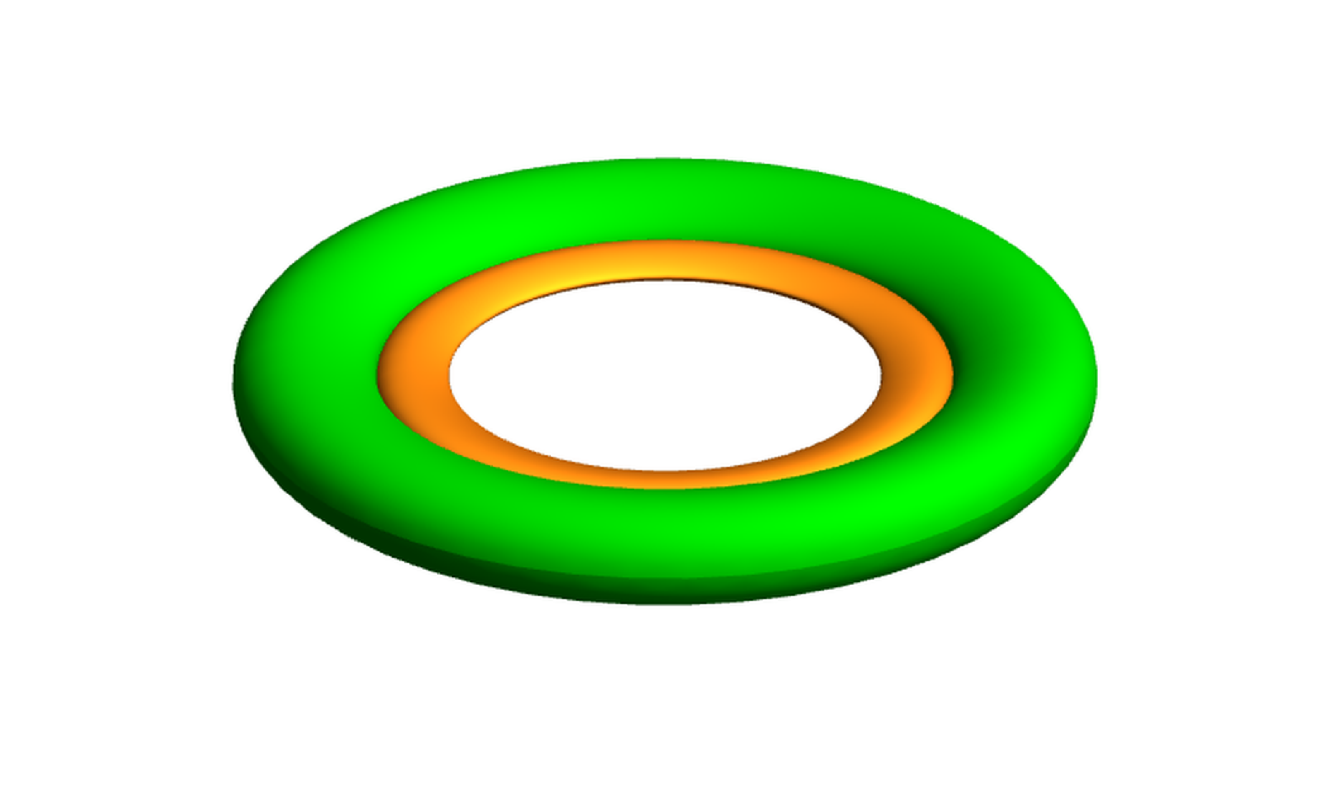}&
\includegraphics[scale=0.3]{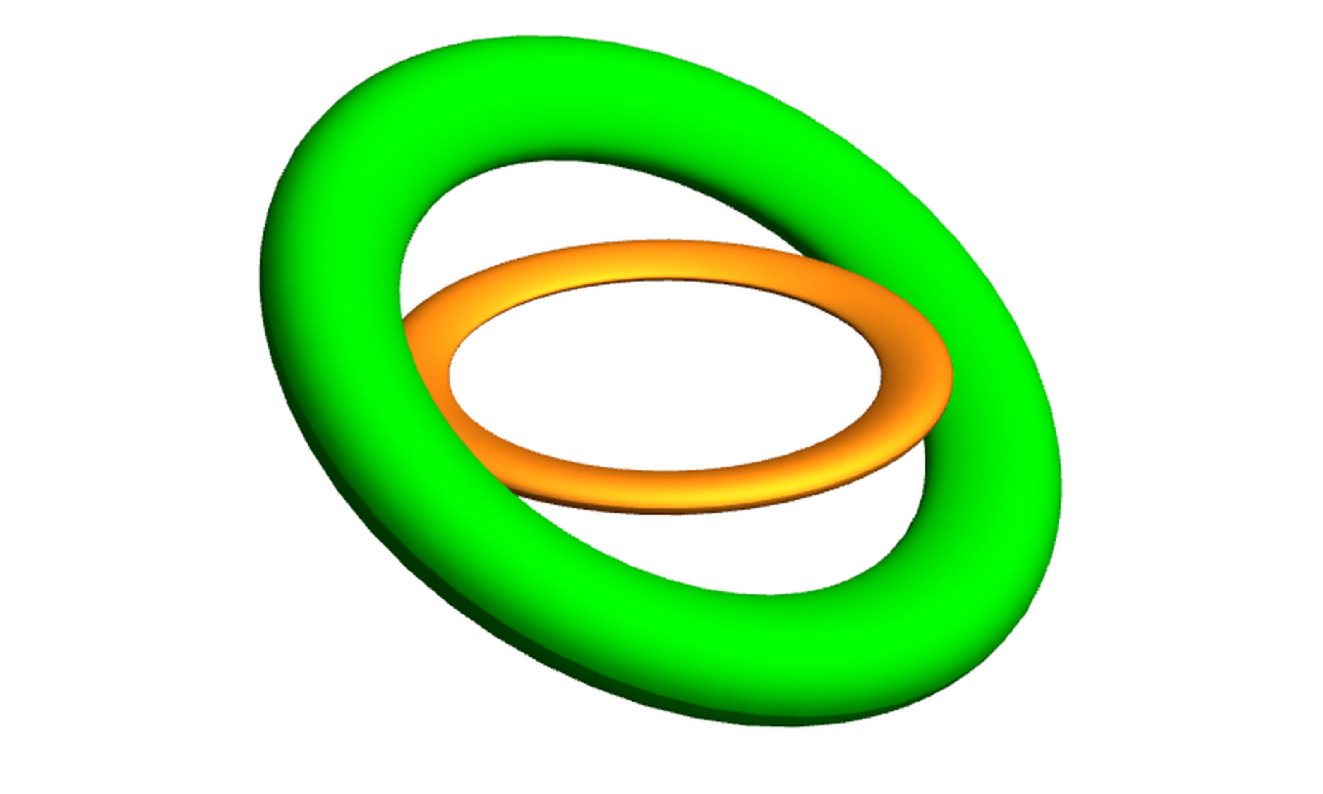}
\tabularnewline
\end{tabular}
\par\end{centering}
\caption{{Three-dimensional visualization of two toroidal
structures. The left panel shows a configuration with parallel symmetry
axes, while the right panel shows a configuration with mutually inclined
symmetry axes.}}
\label{fig:torus3D}
\end{figure}

{
For the \(i\)-th torus we denote by \(v_{(i)}^\mu\) the spacelike unit
vector along its symmetry axis. The inclination angle \(\theta_{(i)}\)
between this axis and the normal \(n^\mu\) to the original equatorial plane
is defined by
}
\begin{equation}
    \cos\theta_{(i)}
    =
    n_\mu v_{(i)}^\mu .
    \label{eq:torus-inclination}
\end{equation}
{This definition allows us to distinguish between configurations with
parallel toroidal symmetry axes and configurations in which the axes are
mutually inclined. Examples of both cases are shown in Fig.~\ref{fig:torus3D}.
Thus, each torus has its own source-adapted polar coordinate
\(\theta'_{(i)}\), which enters the potential
\(W(r,\theta'_{(i)})\). The explicit coordinate transformation and the
transformation of the photon wave vector are not repeated here; they are
given in Appendix~\ref{app:rotated-frame}.
}

{
The ray-tracing procedure is performed backward from the observer's image
plane. For each point \((\alpha,\beta)\) we integrate the corresponding
null geodesic and search for intersections with the surfaces of both tori,
defined by \(W(r,\theta'_{(i)})=W_{0(i)}\). If the ray intersects both
surfaces, the first intersection along the backward ray is used. This
prescription accounts for the geometrical occultation of the more distant
emitting surface by the closer one, but it does not include absorption or
scattering inside the toroidal matter.
}

{
At the emission point the observed-to-emitted frequency ratio is computed
from
\[
    g=\frac{\nu_o}{\nu_e}
    =
    \frac{(k_\mu u^\mu)_o}{(k_\mu u^\mu)_e}.
\]
For a circularly rotating perfect-fluid torus this gives the redshift
formula stated in Appendix~\ref{app:torus-redshift}. The angular velocity
of the fluid is determined by \(l_{\rm disk}\), and the photon angular
momentum entering the redshift expression is evaluated in the rotated
source-adapted frame.
}

{
The bolometric flux map on the observer's screen is constructed using
Liouville's theorem, as discussed in \cite{Sche-Stu:2013:JCAP}
}
\begin{equation}
    F_o(\alpha_i,\beta_j)
    =
    g_{(ij)}^4 I_{e(ij)} ,
    \label{eq:torus-bolometric-flux}
\end{equation}
{
where the factor \(g^4\) corresponds to the transformation of bolometric
intensity. The spectral line profile is computed as
}
\begin{equation}
    F_{\nu_o}
    =
    \sum_{i=1}^{N}
    g_{(i)}^3 I_{\nu_e(i)} ,
    \label{eq:torus-line-profile}
\end{equation}
{
with the local Gaussian emissivity
}
\begin{equation}
    I_{\nu_e(i)}
    =
    \epsilon_0
    \exp
    \left[
    -
    \frac{1}{\sigma^2}
    \left(
    \frac{\nu_o}{g_{(i)}}-\nu_0
    \right)^2
    \right].
    \label{eq:torus-emissivity}
\end{equation}
{
Equations~(\ref{eq:torus-bolometric-flux})--(\ref{eq:torus-emissivity})
are used to simulate the observational signatures of the double-torus
structure in both the bolometric flux distribution
\(F_o(\alpha,\beta)\) and the spectral line profile \(F_{\nu_o}\).
}

\section{Results}

{
We present the results in two steps. First, we analyze an idealized
double-ring system. This simplified model allows us to isolate how the
presence of two emitting radii and two different orbital planes affects
the spectral line profile. Second, we consider finite-thickness
double-torus configurations and study their bolometric flux maps and
one-dimensional $\alpha$-profiles on the observer's screen.
For compactness, we denote the inclination angles of the inner and outer
components by $\theta_A\equiv\theta_{i(A)}$ and
$\theta_B\equiv\theta_{i(B)}$, respectively.
}

\subsection*{Double-ring system}

\begin{figure}[H]
    \centering
    \begin{tabular}{cc}
         \includegraphics[scale=0.4]{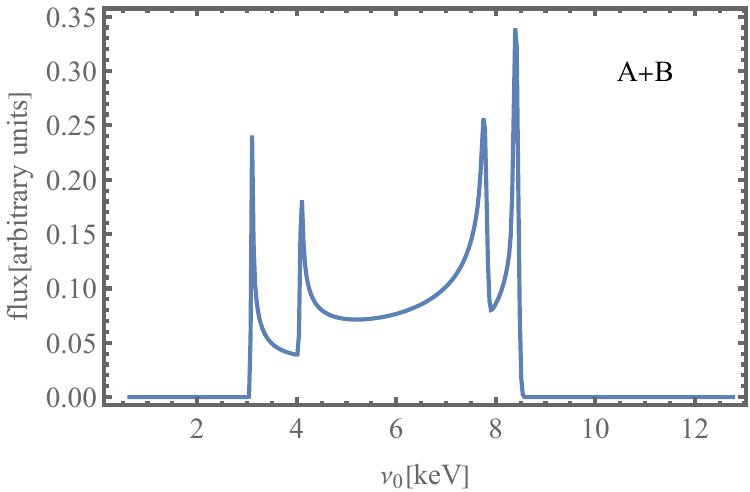}&
         \includegraphics[scale=0.4]{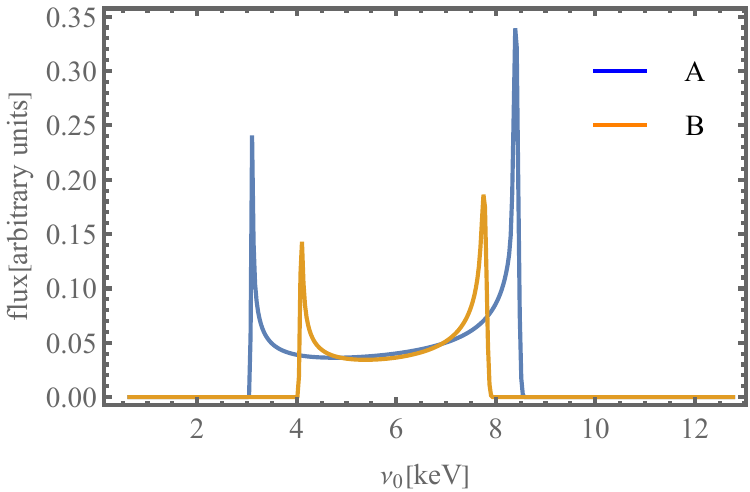}
         \end{tabular}
    \caption{
    Left: spectral line profile of the combined double-ring system
    $A+B$. Right: individual contributions from rings $A$ and $B$.
    The inner ring is fixed at $R_A=6M$ and $\theta_A=0^\circ$,
    while the outer ring in this example has $R_B=10M$ and
    $\theta_B=15^\circ$.
    }
    \label{fig:plring1}
\end{figure}

{
For the double-ring configurations we keep the inner ring $A$ fixed at
$(R_A,\theta_A)=(6M,0^\circ)$. The outer ring $B$ is then varied in the
parameter range
\[
    (R_B,\theta_B)\in[8M,20M]\times[0^\circ,45^\circ].
\]
The observed line profile of the full system is a
superposition of the individual contributions from rings $A$ and $B$.
An illustrative example is shown in Fig.~\ref{fig:plring1}, where the
combined profile is compared with the separate profiles of the two rings.
}

\begin{figure}[H]
    \centering
    \includegraphics[scale=0.5]{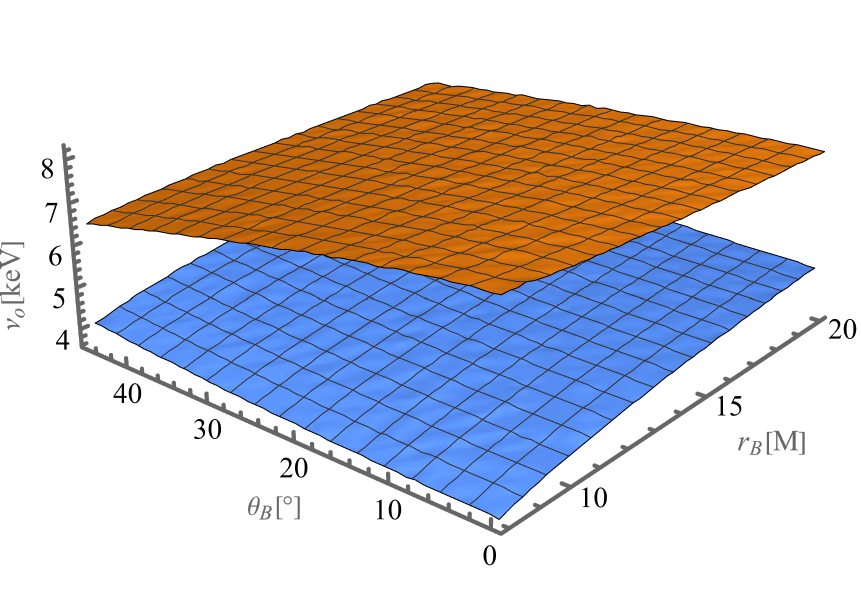}
    \caption{
    Surface plots of the limiting observed frequencies
    $\nu_{o(\min)}$ and $\nu_{o(\max)}$ as functions of the radius
    $R_B$ and inclination $\theta_B$ of the outer ring $B$.
    The inner ring $A$ is kept fixed at the ISCO, $R_A=6M$, in the
    equatorial plane.
    }
    \label{fig:numin_nuomax}
\end{figure}

{
The most important feature of the combined profile is the appearance of a
multi-peak structure. Each ring produces its own pair of redshifted and
blueshifted peaks. If the two components are sufficiently separated in
frequency, the total profile can therefore contain up to four clearly
distinguishable peaks. This feature is a direct imprint of the
multi-component nature of the emitting source.

To quantify how the outer ring affects the line width, we computed the
limiting observed frequencies
\[
    \nu_{o(\min)}=\nu_{o(\min)}(R_B,\theta_B), \qquad
    \nu_{o(\max)}=\nu_{o(\max)}(R_B,\theta_B),
\]
while keeping the inner ring fixed. The resulting surfaces are shown in
Fig.~\ref{fig:numin_nuomax}. They describe how the frequency interval of
the outer component changes with its orbital radius and inclination.
}

\begin{figure}[H]
    \centering
    \includegraphics[scale=0.7]{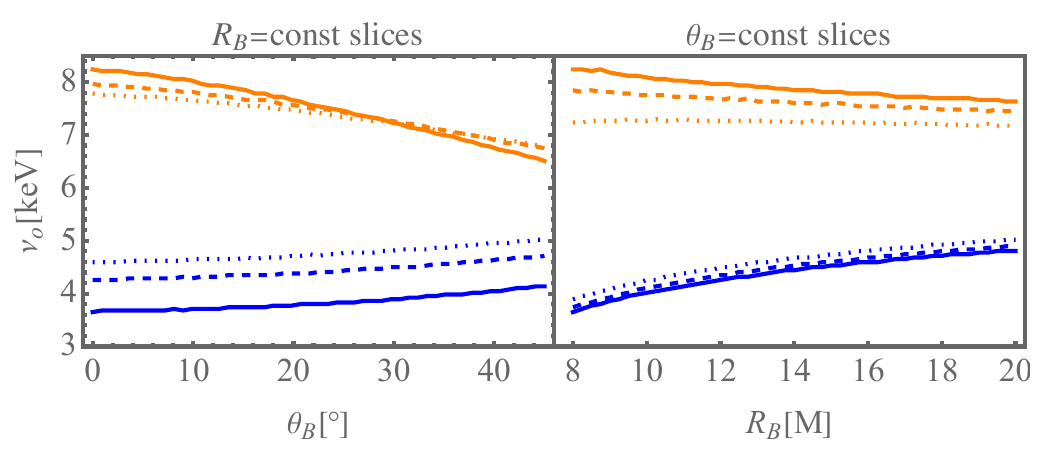}
    \caption{
    Cross-sections of the surfaces shown in
    Fig.~\ref{fig:numin_nuomax}. The left panel shows slices at fixed
    outer-ring radii $R_B=8M$ solid, $12M$ dashed, and $16M$ dotted.
    The right panel shows slices at fixed inclinations
    $\theta_B=0^\circ$ solid, $15.3^\circ$ dashed, and $29.7^\circ$
    dotted.
    }
    \label{fig:numin_nuomax_sect}
\end{figure}

{
The cross-sections in Fig.~\ref{fig:numin_nuomax_sect} show the same
dependence more clearly. Increasing $R_B$ reduces the Keplerian orbital
velocity of the outer ring and therefore narrows the corresponding
frequency interval. The inclination $\theta_B$ changes the projection of
the orbital velocity along the line of sight. In the considered
configuration this also reduces the separation between the limiting
frequencies of the outer-ring contribution. Thus, both the radial position
and the mutual inclination of the rings are directly encoded in the width
and morphology of the observed spectral line.
}

\subsection*{Double-torus system}

{
We now replace the infinitesimally thin rings by finite-thickness
toroidal structures. The tori are described by the Polish-doughnut model
with constant specific angular momentum, $l_{\rm disk}=\mathrm{const}$.
The parameters of the inner torus $A$ and the outer torus $B$ used in the
simulations are listed in Table~\ref{tab:Inner-and-outer}.
}

\begin{table}[H]
\begin{centering}
\begin{tabular}{|c|c|c|c|c|}
\hline 
 & $l_{disk}$ & $W_{0}$ & $r_{in}$ & $r_{out}$\tabularnewline
\hline 
\hline 
$A$ & $3.70328$ & $-0.056239$ & $6.0$ & $8.0$\tabularnewline
\hline 
$B$ & $3.91918$ & $-0.044616$ & $8.0$ & $12.0$\tabularnewline
\hline 
\end{tabular}
\par\end{centering}
\caption{
Parameters of the inner torus $A$ and the outer torus $B$.
For each torus, the values of $l_{\rm disk}$ and $W_0$ are determined
from the conditions
$W_0=W(r_{\rm in},l_{\rm disk})$ and
$W_0=W(r_{\rm out},l_{\rm disk})$.
}\label{tab:Inner-and-outer}
\end{table}

{
To study the effect of mutual misalignment, we consider the following
inclination pattern for the symmetry axes of the two tori:
}

\begin{equation}
(\theta_{i(A)},\theta_{i(B)})=
\left\{
(0^{\circ},0^{\circ}),
(0^{\circ},30^{\circ}),
(0^{\circ},45^{\circ}),
(30^{\circ},0^{\circ}),
(45^{\circ},0^{\circ})
\right\}.
\label{thetasetup}
\end{equation}

{
For each inclination setup we also vary the relative sense of rotation
according to
}

\begin{equation}
(s_{(A)},s_{(B)})=
\left\{
(1,1), (1,-1),(-1,1)
\right\},
\label{spinsetup}
\end{equation}

{
where the signs $s_{(A)}$ and $s_{(B)}$ determine the sign of the
specific angular momentum of each torus,
\[
    l_{{\rm disk}(A)}=s_{(A)}|l_{{\rm disk}(A)}|,\qquad
    l_{{\rm disk}(B)}=s_{(B)}|l_{{\rm disk}(B)}|.
\]
Thus, the case $(1,1)$ corresponds to co-rotation, whereas $(1,-1)$ and
$(-1,1)$ represent counter-rotating configurations.
}

\begin{figure}[H]
\begin{centering}
\includegraphics[scale=0.25]{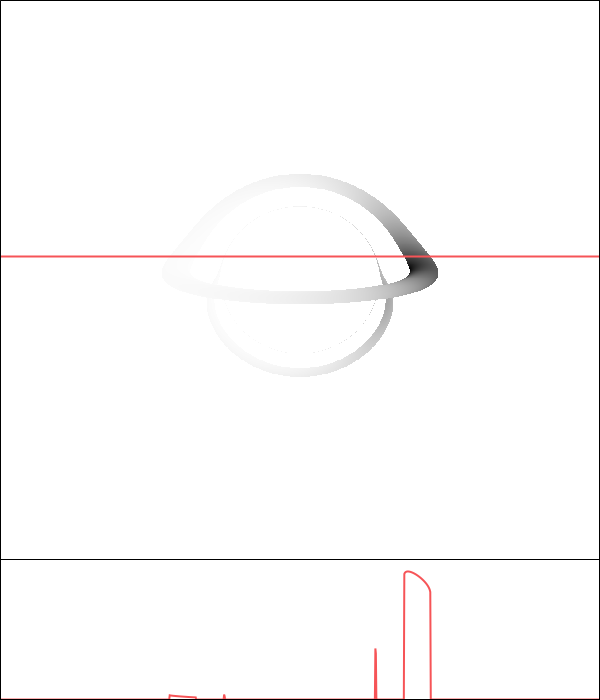}
\par\end{centering}
\caption{{Reference single-torus model. The upper panel shows the bolometric flux
map on the observer's screen, and the lower panel shows the corresponding
$\alpha$-profile extracted at $\beta=1.67$, indicated by the horizontal
red line. The torus is equatorial, $\theta_A=0^\circ$, and rotates with
$s_{(A)}=1$.
}}
\label{fig:single-torus-bolometric}
\end{figure}

\begin{figure}[H]
\begin{centering}
\includegraphics[scale=0.3]{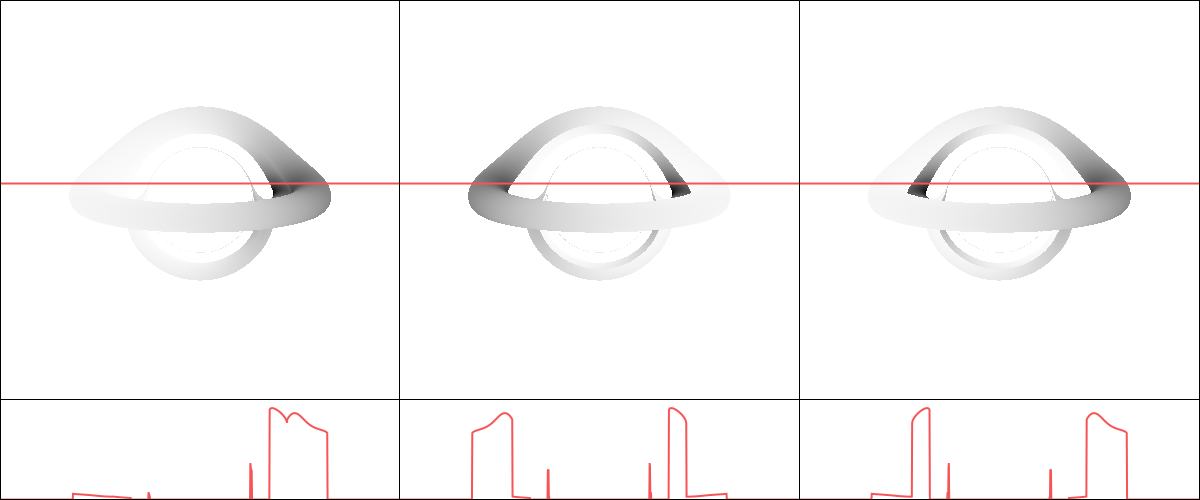}
\par\end{centering}
\caption{{Double-torus bolometric flux maps and corresponding $\alpha$-profiles
for the aligned configuration
$(\theta_A,\theta_B)=(0^\circ,0^\circ)$.
The three columns correspond to
$(s_{(A)},s_{(B)})=(1,1)$ left,
$(1,-1)$ middle, and $(-1,1)$ right.
}}
\label{fig:Double-torus-bolometric1}
\end{figure}

\begin{figure}[H]
\begin{centering}
\includegraphics[scale=0.3]{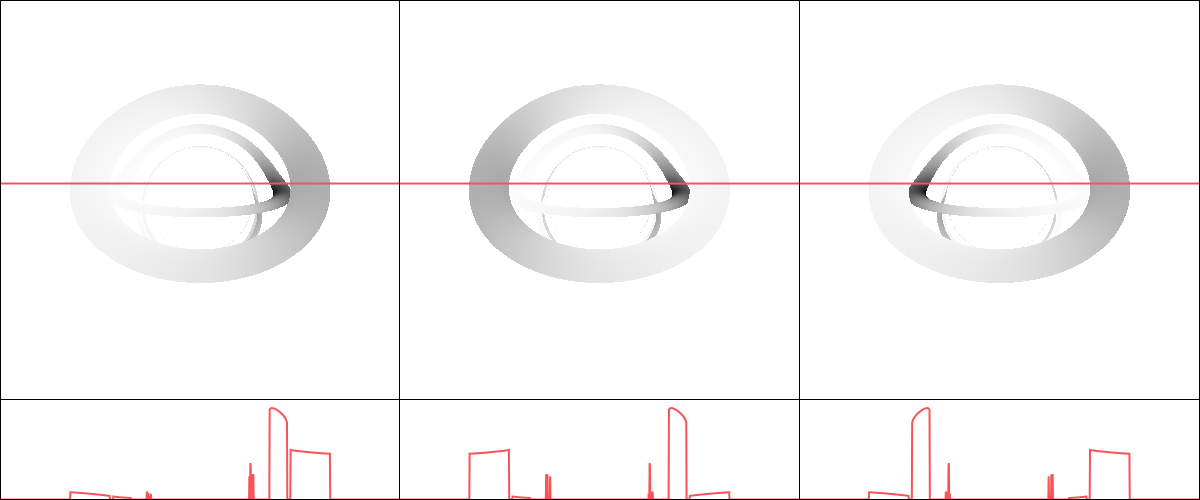}
\par\end{centering}
\caption{{Double-torus bolometric flux maps and corresponding $\alpha$-profiles
for the configuration
$(\theta_A,\theta_B)=(0^\circ,30^\circ)$.
The three columns correspond to
$(s_{(A)},s_{(B)})=(1,1)$ left,
$(1,-1)$ middle, and $(-1,1)$ right.
}}
\label{fig:Double-torus-bolometric2}
\end{figure}

\begin{figure}[H]
\begin{centering}
\includegraphics[scale=0.3]{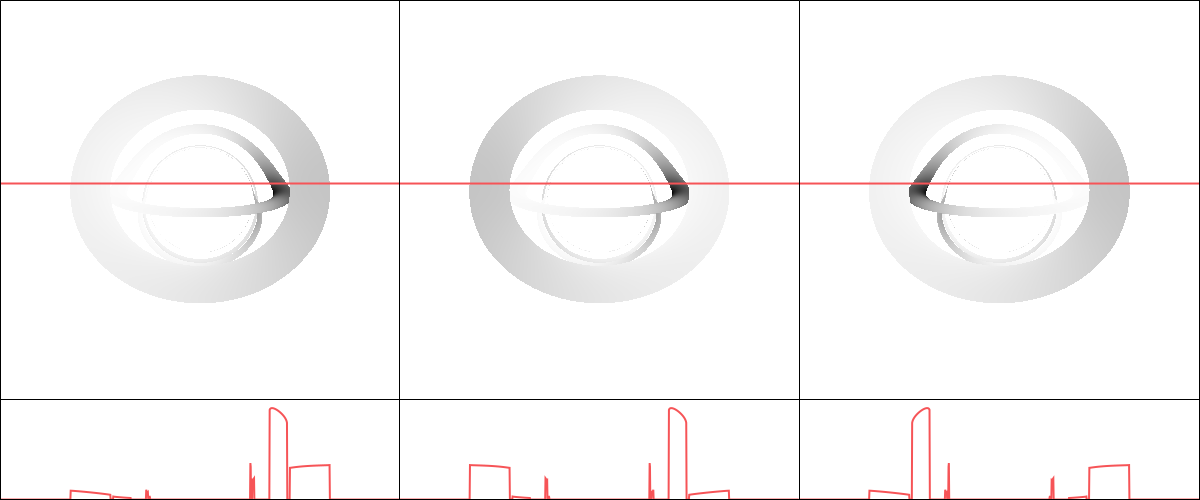}
\par\end{centering}
\caption{{Double-torus bolometric flux maps and corresponding $\alpha$-profiles
for the configuration
$(\theta_A,\theta_B)=(0^\circ,45^\circ)$.
The three columns correspond to
$(s_{(A)},s_{(B)})=(1,1)$ left,
$(1,-1)$ middle, and $(-1,1)$ right.
}}
\label{fig:Double-torus-bolometric3}
\end{figure}

\begin{figure}[H]
\begin{centering}
\includegraphics[scale=0.3]{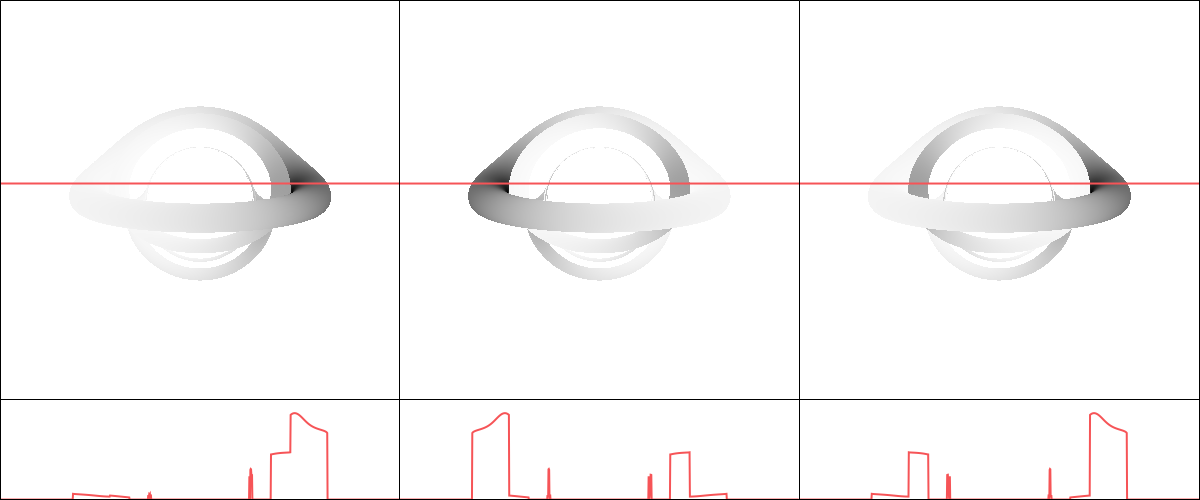}
\par\end{centering}
\caption{{Double-torus bolometric flux maps and corresponding $\alpha$-profiles
for the configuration
$(\theta_A,\theta_B)=(30^\circ,0^\circ)$.
The three columns correspond to
$(s_{(A)},s_{(B)})=(1,1)$ left,
$(1,-1)$ middle, and $(-1,1)$ right.
}}
\label{fig:Double-torus-bolometric4}
\end{figure}

\begin{figure}[H]
\begin{centering}
\includegraphics[scale=0.3]{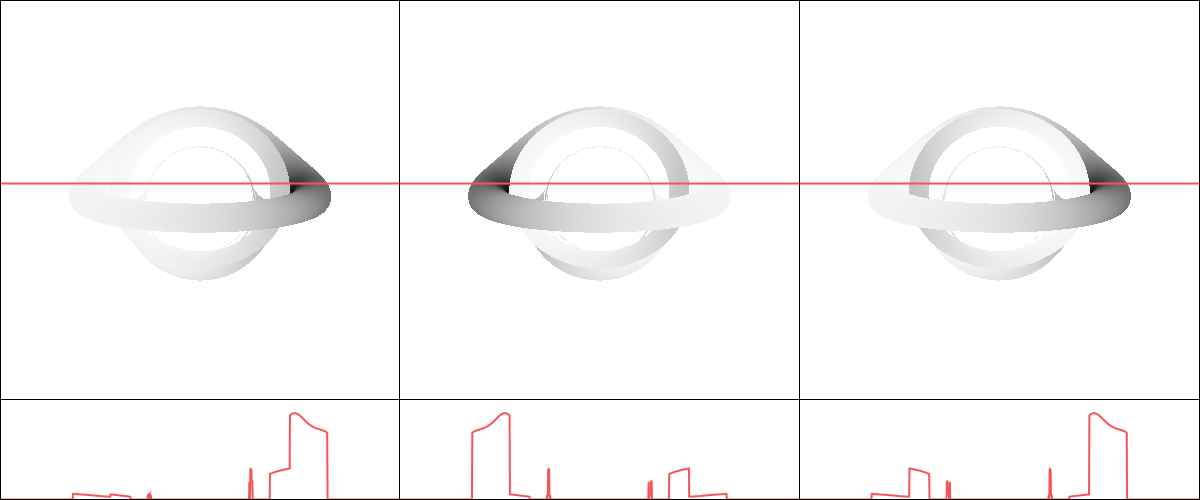}
\par\end{centering}
\caption{{Double-torus bolometric flux maps and corresponding $\alpha$-profiles
for the configuration
$(\theta_A,\theta_B)=(45^\circ,0^\circ)$.
The three columns correspond to
$(s_{(A)},s_{(B)})=(1,1)$ left,
$(1,-1)$ middle, and $(-1,1)$ right.
}}
\label{fig:Double-torus-bolometric5}
\end{figure}

{
For each configuration we compute the bolometric flux map on the
observer's screen and the corresponding one-dimensional
$\alpha$-profile at fixed $\beta=1.67$. The horizontal red line in each
map indicates the value of $\beta$ along which the profile is extracted.
As a reference, Fig.~\ref{fig:single-torus-bolometric} shows the
bolometric flux map and $\alpha$-profile of a single equatorial torus.
The double-torus configurations are shown in
Figs.~\ref{fig:Double-torus-bolometric1}--
\ref{fig:Double-torus-bolometric5}.
}

%frequency distribution assuming  line emission model in the form of Gaussian bell, i.e.  with the emissivity (\ref{emissivity}) with emission parameters  parameters  set to $\sigma=2.5\times10^{-2}$
%, $\nu_{0}=1$, and $\epsilon_{0(A)}=2\epsilon_{0(B)}=1$. The profiled lines are normalized by the profiled line of $(\theta_{i(A)},\theta_{i(B)})=(0^{\circ},0^{\circ})$ case.

{
The reference single-torus image in
Fig.~\ref{fig:single-torus-bolometric} shows the expected asymmetric
flux distribution caused mainly by relativistic Doppler boosting. In the
aligned double-torus case, shown in
Fig.~\ref{fig:Double-torus-bolometric1}, the co-rotating configuration
$(s_{(A)},s_{(B)})=(1,1)$ produces a single dominant Doppler-enhanced
region in the $\alpha$-profile. By contrast, the counter-rotating
configurations $(1,-1)$ and $(-1,1)$ produce two dominant flux peaks,
because the approaching sides of the two tori appear on opposite sides of
the observer's image.
}

{
The same distinction between co-rotating and counter-rotating systems is
preserved when one of the tori is inclined. However, the mutual
misalignment changes the projected velocity field and the visible area of
each emitting component. As a result, the relative heights and widths of
the Doppler peaks in the $\alpha$-profiles are modified. The effect is
particularly clear when comparing configurations in which the outer torus
is tilted,
Figs.~\ref{fig:Double-torus-bolometric2} and
\ref{fig:Double-torus-bolometric3}, with configurations in which the
inner torus is tilted,
Figs.~\ref{fig:Double-torus-bolometric4} and
\ref{fig:Double-torus-bolometric5}. Therefore, the $\alpha$-profile
provides a compact diagnostic of both the relative rotation sense and the
mutual inclination of the two toroidal components.
}

\section{{Discussion}}

{
We have studied the ray-traced observational appearance of idealized
double-ring and double-torus emitting structures in Schwarzschild
spacetime. The main purpose of the model is not to describe a fully
self-consistent dynamical equilibrium, but to isolate the effect of a
non-coplanar multi-component source on spectral line profiles and
bolometric flux distributions. In this sense, the Schwarzschild geometry
provides a useful reference case, because the spherical symmetry of the
background allows us to change the orientation of the emitting structures
without introducing frame-dragging effects. The differences between the
models can therefore be attributed directly to the geometry and motion of
the source.
}

{
The double-ring model shows that a multi-component source can naturally
produce a multi-peak spectral line profile. Each radiating ring gives rise
to its own pair of redshifted and blueshifted peaks, and the total profile
is their superposition. The appearance of up to four distinguishable peaks
therefore does not require a complicated emissivity prescription; it can
already arise from two narrow emitting components located at different
radii and, in general, in different orbital planes. The radius of the
outer ring controls the local Keplerian velocity and hence the width of
the corresponding component of the line profile. The inclination of the
outer ring changes the projection of the orbital velocity along the line
of sight and modifies the separation and relative prominence of the
redshifted and blueshifted peaks.
}

{
The finite-thickness double-torus model demonstrates that the same basic
geometrical effects are also visible in bolometric flux maps. In this
case, the observed radiation is affected not only by gravitational
redshift and Doppler boosting, but also by the finite size and projected
shape of the emitting surfaces. The relative sense of rotation of the two
tori has a particularly clear observational imprint. In co-rotating
configurations, the Doppler-enhanced regions of the two components tend
to appear on the same side of the image and can merge into a single
dominant maximum in the $\alpha$-profile. In counter-rotating
configurations, the approaching sides of the two tori are located on
opposite sides of the image, producing two dominant peaks in the
bolometric-flux profile.
}

{
The mutual misalignment of the toroidal axes introduces an additional
asymmetry. It changes the visible projected area of each torus and the
line-of-sight component of its orbital velocity. As a result, the
misalignment is imprinted mainly in the amplitudes, widths, and relative
positions of the Doppler-enhanced peaks. The $\alpha$-profile is therefore
a compact diagnostic tool: it preserves information about both the
relative rotation sense and the mutual inclination of the emitting
components.
}  

{
The present study should be understood as a simplified radiative
experiment. The two tori are prescribed as stationary Polish-doughnut
surfaces, and the double structure is treated as an instantaneous
radiating configuration. We do not claim that the full non-coplanar
double-torus system represents a permanently stable global hydrodynamical
equilibrium. Possible dynamical interaction between the tori, mass
exchange, turbulence, magnetic fields, self-gravity of the fluid,
absorption, scattering, and time-dependent radiative transfer are not
included. These effects can be important in realistic accretion flows, but
their omission allows us to identify the basic geometrical signatures of
non-coplanar double structures in a transparent way.
}

{
The results can therefore be used as reference patterns for more
sophisticated studies. In particular, the multi-peak line profiles and the
one- or two-peak structure of the bolometric $\alpha$-profiles provide
simple signatures against which more realistic GRMHD and radiative-transfer
models of tilted, warped, or interacting accretion flows may be compared.
A natural continuation of this work would be to extend the model to Kerr
spacetime, include time-dependent evolution of interacting tori, and study
the role of absorption and emission inside the toroidal matter.
}

\section{{Conclusions}}

{
We constructed ray-traced models of idealized double-ring and double-torus
radiating structures orbiting a Schwarzschild black hole. The two emitting
components were allowed to have mutually inclined symmetry axes. The main
results can be summarized as follows.
}

\begin{enumerate}

\item {
A double-ring source produces a spectral line profile that is a
superposition of the individual profiles of the two rings. Since each ring
generates its own redshifted and blueshifted peak, the combined profile
can contain up to four distinguishable peaks.
}

\item {
The radius of the outer ring and its inclination with respect to the inner
ring are encoded in the width and morphology of the spectral line profile.
Increasing the outer-ring radius reduces the orbital velocity and narrows
the corresponding frequency interval, while changing the inclination
modifies the line-of-sight velocity projection.
}

\item {
For finite-thickness double-torus configurations, the relative sense of
rotation has a clear imprint on the bolometric flux maps and
$\alpha$-profiles. Co-rotating tori tend to produce one dominant
Doppler-enhanced region, whereas counter-rotating configurations can
produce two dominant flux peaks on opposite sides of the image.
}

\item {
The mutual misalignment of the toroidal axes affects the relative
amplitudes, widths, and positions of the Doppler-enhanced peaks. Thus, the
$\alpha$-profile of the bolometric flux provides a simple diagnostic of
both the orientation and the rotation sense of the components.
}

\item {
Although the model is idealized, it shows that non-coplanar
multi-component accretion structures can leave characteristic signatures
in spectral line profiles and bolometric-flux distributions. These
signatures may be useful for interpreting more realistic simulations of
tilted, warped, or interacting accretion flows.
}

\end{enumerate}

\section*{Acknowledgment}
This research was supported by the Research Centre for Theoretical Physics and Astrophysics, Institute of Physics, Silesian University in Opava, and by the Silesian University in Opava under grant No. SGS/24/2024. D.O. also acknowledges support from the grant No. IGS/22/2026.

\appendix

\section{Ray tracing and coordinate transformations}
\label{app:raytracing}

In this appendix we summarize the ray-tracing equations and the coordinate
transformations used in the construction of the ring and torus images. We
use geometrized units \(G=c=M=1\).

\subsection{Null geodesics in Schwarzschild spacetime}
\label{app:null-geodesics}

The Schwarzschild line element is written in the standard form
\begin{equation}
    ds^2
    =
    -\left(1-\frac{2}{r}\right)dt^2
    +
    \left(1-\frac{2}{r}\right)^{-1}dr^2
    +
    r^2\left(d\theta^2+\sin^2\theta\,d\phi^2\right).
\end{equation}
For null geodesics, the motion is characterized by the conserved energy
\(E=-k_t\), the axial angular momentum \(L=k_\phi\), and the Carter
constant \(Q\). We introduce the impact parameters
\begin{equation}
    l=\frac{L}{E},
    \qquad
    q=\frac{Q}{E^2}.
\end{equation}
It is useful to use the variables
\begin{equation}
    u=\frac{1}{r},
    \qquad
    m=\cos\theta .
\end{equation}
The separated radial and latitudinal potentials are
\begin{equation}
    U(u;l,q)
    =
    1-(l^2+q)u^2+2(l^2+q)u^3 ,
    \label{eq:app-U}
\end{equation}
and
\begin{equation}
    M(m;l,q)
    =
    q-(l^2+q)m^2 .
    \label{eq:app-M}
\end{equation}

For a ring whose orbital plane is inclined with respect to the observer's
reference plane, we introduce a coordinate system adapted to the ring. In
this system the emitting ring lies in the plane \(m_e=0\). The observer's
latitudinal coordinate in the ring-adapted frame is denoted by
\begin{equation}
    \bar m_o=\cos(\theta_o-\theta_i),
    \label{eq:app-mo-bar}
\end{equation}
where \(\theta_i\) is the inclination of the ring plane. Equivalently, if
\(n_i^\mu\) is the unit normal to the \(i\)-th ring plane, the inclination
can be defined by
\begin{equation}
    \cos\theta_i
    =
    \tilde{\omega}^{(3)}_{\mu} n_i^\mu ,
    \label{eq:app-ring-inclination}
\end{equation}
where \(\tilde{\omega}^{(3)}_{\mu}\) denotes the spatial tetrad direction
normal to the observer's reference equatorial plane.

For a photon emitted from the ring and received by the observer, the
radial and latitudinal parts of the null geodesic are related by Carter's
integral equation
\begin{equation}
    \int\limits_{u_o}^{u_e}
    \frac{\diff u}{\sqrt{U(u;l,q)}}
    -
    m_{\rm sgn}
    \int\limits_{\bar m_o}^{0}
    \frac{\diff m}{\sqrt{M(m;l,q)}}
    =
    0 .
    \label{eq:app-carter-direct}
\end{equation}
Here \(u_o\) and \(u_e\) are the inverse radial coordinates of the
observer and emitter, respectively. The factor \(m_{\rm sgn}=\pm1\)
specifies the initial direction of the latitudinal motion. Equation
(\ref{eq:app-carter-direct}) is the form used to determine the impact
parameters of photon trajectories connecting the observer with a point on
the emitting ring when no additional turning-point contribution is written
explicitly.

More generally, if the photon trajectory contains radial or latitudinal
turning points, the integrals in Eq.~(\ref{eq:app-carter-direct}) are
understood as sums over the corresponding monotonic parts of the motion.
We denote this generalized form as
\begin{equation}
    \mathcal{I}_u(u_o,u_e;n_u)
    -
    m_{\rm sgn}
    \mathcal{I}_m(\bar m_o,0;n_m)
    =
    0 .
    \label{eq:app-carter-general}
\end{equation}
The integers \(n_u\) and \(n_m\) count the number of radial and
latitudinal turning points along the photon trajectory. The turning points
are determined by the roots of
\begin{equation}
    U(u;l,q)=0,
    \qquad
    M(m;l,q)=0.
\end{equation}
In the present work we include the primary image contributions specified
by
\begin{equation}
    (n_u,n_m)
    =
    \{(0,0),(0,1),(1,1)\}.
    \label{eq:app-image-orders}
\end{equation}
The contribution \((0,0)\) corresponds to a direct trajectory without
turning points, \((0,1)\) includes one latitudinal turning point, and
\((1,1)\) includes one radial and one latitudinal turning point. In the
numerical implementation the relevant turning-point contributions are
added to the radial and latitudinal integrals in
Eq.~(\ref{eq:app-carter-general}).

The observer's screen coordinates \((\alpha,\beta)\) are related to the
impact parameters by
\begin{equation}
    l=\alpha\sqrt{1-m_o^2},
    \qquad
    q=\beta^2+m_o^2\alpha^2,
    \label{eq:app-impact}
\end{equation}
where \(m_o=\cos\theta_o\) is the observer's latitudinal coordinate in the
global Schwarzschild frame. Each point on the observer's image plane
therefore defines one null geodesic through the parameters \(l\) and
\(q\). For the inclined-ring calculation, the same geodesic is evaluated
in the ring-adapted frame, where the observer coordinate entering
Eq.~(\ref{eq:app-carter-direct}) is \(\bar m_o\).

\subsection{Spectral line profile of a ring}
\label{app:ring-line}

For the ring model, the observed specific flux is computed by integrating
over the redshift factor \(g=\nu_o/\nu_e\). For a narrow emitting ring at
radius \(r_e\), we use
\begin{equation}
    F(\nu_o;r_e)
    =
    \Delta r_e
    \int_{g_{\min}}^{g_{\max}}
    g^3
    I_e(\nu_o/g;r_e)
    \left|
    \frac{\partial(\alpha,\beta)}{\partial(r_e,g)}
    \right|
    dg .
    \label{eq:app-ring-flux}
\end{equation}
The factor \(g^3\) follows from the invariance of \(I_\nu/\nu^3\) along a
null geodesic. The Jacobian
\begin{equation}
    \left|
    \frac{\partial(\alpha,\beta)}{\partial(r_e,g)}
    \right|
\end{equation}
is evaluated numerically from the ray-tracing map
\((\alpha,\beta)\mapsto(r_e,g)\). Equivalently,
\begin{equation}
    \left|
    \frac{\partial(\alpha,\beta)}{\partial(r_e,g)}
    \right|
    =
    \left|
    \frac{\partial(r_e,g)}{\partial(\alpha,\beta)}
    \right|^{-1}.
\end{equation}

\subsection{Rotated coordinate frame for an inclined torus}
\label{app:rotated-frame}

For a torus whose symmetry axis is inclined with respect to the original
equatorial axis, we introduce a source-adapted coordinate system
\((r',\theta',\phi')\). The primed frame is obtained by a rotation of the
source-adapted spatial triad about the original \(y\)-axis by the angle
\(\theta_i\). Equivalently, the new symmetry axis lies in the original
\(x-z\) plane, with \(x\)-axis directed towards the observer, and has the direction
\begin{equation}
    \mathbf{e}_{z'}
    =
    \sin\theta_i\,\mathbf{e}_x
    +
    \cos\theta_i\,\mathbf{e}_z .
\end{equation}
The remaining basis vectors may be chosen as
\begin{equation}
    \mathbf{e}_{x'}
    =
    \cos\theta_i\,\mathbf{e}_x
    -
    \sin\theta_i\,\mathbf{e}_z,
    \qquad
    \mathbf{e}_{y'}=\mathbf{e}_y .
\end{equation}
With this convention, the transformation between the original spherical
coordinates \((r,\theta,\phi)\) and the rotated coordinates
\((r',\theta',\phi')\) is
\begin{equation}
    r'=r,
\end{equation}
\begin{equation}
    \cos\theta'
    =
    \sin\theta\cos\phi\sin\theta_i
    +
    \cos\theta\cos\theta_i ,
    \label{eq:app-costhetap}
\end{equation}
and
\begin{equation}
    \tan\phi'
    =
    \frac{
    \sin\theta\sin\phi
    }{
    \sin\theta\cos\phi\cos\theta_i
    -
    \cos\theta\sin\theta_i
    } .
    \label{eq:app-tanphip}
\end{equation}
In numerical calculations it is preferable to use the two-argument
function
\begin{equation}
    \phi'
    =
    {\rm atan2}
    \left(
    \sin\theta\sin\phi,\,
    \sin\theta\cos\phi\cos\theta_i
    -
    \cos\theta\sin\theta_i
    \right),
\end{equation}
which fixes the correct quadrant of \(\phi'\).

The components of a photon wave vector
\begin{equation}
    k^\mu=(k^t,k^r,k^\theta,k^\phi)
\end{equation}
transform to the primed frame according to
\begin{equation}
    k^{\mu'}
    =
    \frac{\partial x^{\mu'}}{\partial x^\nu}k^\nu .
    \label{eq:app-vector-transform}
\end{equation}
The non-zero elements required in the present calculation are
\begin{equation}
    \frac{\partial t'}{\partial t}=1,
    \qquad
    \frac{\partial r'}{\partial r}=1,
\end{equation}
\begin{equation}
    \frac{\partial\theta'}{\partial\theta}
    =
    -
    \frac{
    \cos\theta\cos\phi\sin\theta_i
    -
    \sin\theta\cos\theta_i
    }{
    \sqrt{1-
    \left(
    \sin\theta\cos\phi\sin\theta_i
    +
    \cos\theta\cos\theta_i
    \right)^2}
    },
\end{equation}
\begin{equation}
    \frac{\partial\theta'}{\partial\phi}
    =
    \frac{
    \sin\theta\sin\phi\sin\theta_i
    }{
    \sqrt{1-
    \left(
    \sin\theta\cos\phi\sin\theta_i
    +
    \cos\theta\cos\theta_i
    \right)^2}
    }.
\end{equation}
It is convenient to define
\begin{equation}
    X
    =
    \sin\theta\cos\phi\cos\theta_i
    -
    \cos\theta\sin\theta_i,
    \qquad
    Y
    =
    \sin\theta\sin\phi,
\end{equation}
so that
\begin{equation}
    \phi'={\rm atan2}(Y,X),
    \qquad
    X^2+Y^2=\sin^2\theta' .
\end{equation}
Then
\begin{equation}
    \frac{\partial\phi'}{\partial\theta}
    =
    -
    \frac{
    \sin\phi\sin\theta_i
    }{
    X^2+Y^2
    },
\end{equation}
and
\begin{equation}
    \frac{\partial\phi'}{\partial\phi}
    =
    \frac{
    \sin^2\theta\cos\theta_i
    -
    \sin\theta\cos\theta\cos\phi\sin\theta_i
    }{
    X^2+Y^2
    }.
\end{equation}
For \(\theta_i=0\), the primed and unprimed frames coincide:
\[
    \theta'=\theta,
    \qquad
    \phi'=\phi.
\]

\subsection{Geodesic integration for the torus images}
\label{app:torus-geodesics}

For the construction of torus images we integrate the geodesic equations
in terms of \(u=1/r\) and \(m=\cos\theta\). The relevant geodesic equations for null vector $k^{\mu}$
in the form \cite{Sche-Stu:2022:JCAP} are
\begin{eqnarray}
    \frac{dk^u}{d\lambda}
    &=&
    2u^3
    \left(
    U+\frac{u}{4}\frac{dU}{du}
    \right),
    \\
    \frac{dk^m}{d\lambda}
    &=&
    \frac{2}{u}k^u k^m
    +
    \frac{1}{2}u^4\frac{dM}{dm},
    \\
    \frac{d\phi}{d\lambda}
    &=&
    \frac{u^2 l}{1-m^2}.
\end{eqnarray}
Here \(U=U(u;l,q)\) and \(M=M(m;l,q)\) are the potentials defined above.
For a given observer position \((u_o,m_o)\) and screen coordinates
\((\alpha,\beta)\), the initial values are
\begin{eqnarray}
    k_o^u
    &=&
    u_o^2
    \sqrt{
    1-u_o^2(l^2+q)(1-2u_o)
    },
    \\
    k_o^m
    &=&
    u_o^2
    \sqrt{
    q-m_o^2(l^2+q)
    },
    \\
    \phi_o&=&0.
\end{eqnarray}
The signs of \(k_o^u\) and \(k_o^m\) are chosen according to the required
direction of backward ray tracing from the observer to the emitting
surface.

\subsection{Intersection with a torus and frequency shift}
\label{app:torus-redshift}

The torus surface is defined by a constant value of the effective
potential,
\begin{equation}
    W(r,\theta')=W_0.
\end{equation}
During the ray-tracing procedure we search for intersections of a null
geodesic with the surfaces of both tori. If a photon intersects both
emitting surfaces, the contribution with the smaller value of the affine
parameter \(\lambda\) is used, corresponding to the first surface reached
by the photon when traced backward from the observer.

The frequency shift is defined as
\begin{equation}
    g
    =
    \frac{\nu_o}{\nu_e}
    =
    \frac{(k_\mu u^\mu)_o}{(k_\mu u^\mu)_e}.
\end{equation}
For a static observer at infinity and a circularly rotating emitter in the
primed torus frame, this gives
\begin{equation}
    g
    =
    \frac{
    \sqrt{
    1-\frac{2}{r}
    -
    r^2\sin^2\theta'\,\Omega^2
    }
    }{
    1-l'\Omega
    }.
    \label{eq:app-redshift}
\end{equation}
The angular velocity of the perfect-fluid torus is
\begin{equation}
    \Omega
    =
    \frac{
    \left(1-\frac{2}{r}\right)l_{\rm disk}
    }{
    r^2\sin^2\theta'
    } .
    \label{eq:app-omega}
\end{equation}
The quantity \(l'\) is the photon impact parameter with respect to the
rotated azimuthal coordinate,
\begin{equation}
    l'
    =
    -\frac{k_{\phi'}}{k_t}.
    \label{eq:app-lprime}
\end{equation}
In the numerical implementation it is evaluated from the transformed
photon momentum \(k^{\mu'}\), rather than from the unprimed value of
\(l\) alone. Explicitly,
\begin{equation}
    k^{\phi'}
    =
    \frac{\partial\phi'}{\partial\theta}k^\theta
    +
    \frac{\partial\phi'}{\partial\phi}k^\phi,
\end{equation}
and hence
\begin{equation}
    l'
    =
    \frac{
    r^2\sin^2\theta'\,k^{\phi'}
    }{
    \left(1-\frac{2}{r}\right)k^t
    }.
    \label{eq:app-lprime-vector}
\end{equation}
This definition is invariantly equivalent to Eq.~(\ref{eq:app-lprime})
and is the safest way to evaluate the redshift for an inclined torus.

\subsection{Bolometric flux and line profile}
\label{app:flux}

Using Liouville's theorem, the bolometric intensity transforms as
\(I_o=g^4 I_e\). Therefore, the bolometric flux map on the observer's
screen is computed as
\begin{equation}
    F_o(\alpha_i,\beta_j)
    =
    g_{(ij)}^4 I_{e(ij)} .
    \label{eq:app-bolometric}
\end{equation}
For spectral line calculations we use the transformation of the specific
intensity, \(I_{\nu_o}=g^3 I_{\nu_e}\), and write
\begin{equation}
    F_{\nu_o}
    =
    \sum_{i=1}^{N}
    g_{(i)}^3 I_{\nu_e(i)} .
    \label{eq:app-line-sum}
\end{equation}
The local emissivity is modeled as a Gaussian line,
\begin{equation}
    I_{\nu_e(i)}
    =
    \epsilon_0
    \exp
    \left[
    -
    \frac{1}{\sigma^2}
    \left(
    \frac{\nu_o}{g_{(i)}}-\nu_0
    \right)^2
    \right].
    \label{eq:app-emissivity}
\end{equation}
Equations~(\ref{eq:app-bolometric})--(\ref{eq:app-emissivity}) are used to
construct both the bolometric flux maps and the spectral line profiles of
the double-torus configurations.

\bibliographystyle{unsrt}
\bibliography{references}

\end{document}